\documentclass[twocolumn,cleanfoot,10pt]{asme2ej}

\usepackage{float}
\usepackage{graphicx}
\usepackage{hyperref}
\usepackage[switch]{lineno}
\usepackage{times}
\usepackage{epsfig}
\usepackage{bm}
\usepackage{breakurl}
\usepackage{graphicx, subfigure}
\usepackage{amsmath,amssymb, amsfonts, euscript, mathrsfs}
\usepackage{subeqnarray}
\usepackage{cases}
\usepackage{enumitem}
\usepackage{algorithm}
\usepackage{algorithmic}
\usepackage{fancybox}
\usepackage[table]{xcolor}
\usepackage{wrapfig}
\usepackage{array}
\usepackage{multirow}
\usepackage{hhline}
\usepackage[numbered,framed]{matlab-prettifier}
\definecolor{lightgray}{gray}{0.9}
\definecolor{lightblue}{rgb}{0.98,0.98,1.0}

\usepackage[normalem]{ulem}
\newcommand{\stkout}[1]{\ifmmode\text{\sout{\ensuremath{#1}}}\else\sout{#1}\fi}

\let\oldequation\equation
\let\oldendequation\endequation

\renewenvironment{equation}
  {\linenomathNonumbers\oldequation}
  {\oldendequation\endlinenomath}

\title{A Streamline-guided De-Homogenization Approach for Structural Design}

\author{Junpeng Wang
    \affiliation{
	Computer Graphics and Visualization\\
	Department of Informatics\\
	Technical University of Munich\\
	Garching 85748, Germany\\
    Email: junpeng.wang@tum.de
    }	
}

\author{R\"udiger Westermann
    \affiliation{
	Computer Graphics and Visualization\\
	Department of Informatics\\
	Technical University of Munich\\
	Garching 85748, Germany\\
    Email: westermann@tum.de
    }	
}

\author{Jun Wu\thanks{Address all correspondence to this author.}
    \affiliation{
	Department of Sustainable \\
	Design Engineering\\
	Delft University of Technology\\
	Delft 2628 CE, The Netherlands\\
    Email: j.wu-1@tudelft.nl
    }	
}

\begin{document}

\maketitle

\begin{abstract}
{\it We present a novel de-homogenization approach for efficient design of high-resolution load-bearing structures. The proposed approach builds upon a streamline-based parametrization of the design domain, using a set of space-filling and evenly-spaced streamlines in the two mutually orthogonal direction fields that are obtained from homogenization-based topology optimization. Streamlines in these fields are converted into a graph, which is then used to construct a quad-dominant mesh whose edges follow the direction fields. In addition, the edge width is adjusted according to the density and anisotropy of the optimized orthotropic cells. In a number of numerical examples, we demonstrate the mechanical performance and regular appearance of the resulting structural designs, and compare them with those from classic and contemporary approaches.
}
\end{abstract}



\section{Introduction}

Achieving the highest stiffness while using the least amount of material is a fundamental task in mechanical design. This is often formulated as an optimization problem, e.g., topology optimization, in which the material distribution is optimized~\cite{Bendsoe2004Book,Sigmund2013SMO}. Early works in topology optimization employ a material model corresponding to infinitely small square cells with rectangular holes~\cite{Bendsoe1988CMAME}. The orientation of the cell and the size of the rectangular hole therein are optimized to minimize the compliance of the structure. The material properties of these orthotropic cells are constructed using homogenization. This homogenization-based approach generates a mathematical specification of theoretically optimal structures. Yet how to translate the specification of these spatially-varying orthotropic cells into a globally consistent geometry has remained a challenge. The lack of a consistent geometry means that the optimal structure is not manufacturable. To circumvent this problem, the focus of research in topology optimization has since the 1990s shifted to optimizing the distribution of solid isotropic materials. Popular approaches such as those based on density~\cite{Bendsoe1989SO,Sigmund2001SMO}, level-sets~\cite{Wang2003CMAME,Allaire2004JCP}, evolutionary procedures~\cite{Xie1993CS}, and explicit geometric descriptions~\cite{Norato2004IJNME,Guo2014JAM}, all belong to this category. 

Recent years have seen a revival of homogenization-based approaches, with a focus on the post-process of translating the results of homogenization-based topology optimization into a manufacturable geometry. This efficiently generates high-resolution structural designs since topology optimization is performed only on a coarse grid. This post-process is now often referred to as \textit{de-homogenization}. Pantz et al. proposed one of the first solutions towards this end~\cite{Pantz2008JCO}, which was revisited and improved by Groen and Sigmund~\cite{Groen2018IJNME} and Allaire et al.~\cite{Allaire2018CMA}. These approaches have since been extended to 3D~\cite{Groen2020CMAME,geoffroy2020}, and to deal with singularities in the optimized orientation fields~\cite{Stutz2020SMO}. A key component in these approaches is computing a fine-grid scalar field whose gradients are aligned with optimized orientations from homogenization-based topology optimization. Wu et al. reformulated this post-process as quad/hex-dominant meshing, i.e., constructing quad/hex-dominant meshes whose edges are aligned with the optimized orientations~\cite{Wu2021TVCG}. Stutz et al.~\cite{stutz2022synthesis} reported a method to generate high-resolution multi-laminar structures from frame fields by tracing the \emph{stream surface}. They further formulated the finding of such a set of well-spaced stream surfaces as an optimization problem. Convolutional neural networks have also been found useful for de-homogenization~\cite{Elingaard2022}. Alternative de-homogenization approaches include~\cite{lee2021design,Zhu2019JMPS}.

In this paper, we propose a streamline-guided de-homogenization approach. Similar to the aforementioned de-homogenization approaches, our approach takes as input the width and orientation of spatially-varying square cells that are optimized via homogenization-based topology optimization. In contrast to the majority of prior de-homogenization approaches that represent the final structure as a binary field, our approach generates an explicit representation in the form of a quad-dominant mesh. The edges of the mesh represent beam-like sub-structures. The edges are aligned with the optimized cell orientations, and each edge is assigned a unique width. This compact representation is beneficial for downstream operations such as user editing and fabrication process planning. 

Our technical contribution is a novel method to convert the result of the homogenization-based optimization process, i.e., the optimized cell widths and orientations, into a domain-filling mesh whose elements are then de-homogenized consistently. Our approach builds upon streamlines, which are commonly used for flow and stress tensor field visualization. Our approach avoids the projection step to optimize for a consistent fine-grid scalar field (e.g., in~\cite{Groen2018IJNME}), and thus is computationally efficient. We first parameterize the design domain using a set of domain-filling and evenly-spaced streamlines that are aligned with the edges of optimized cells. The streamlines are then converted into a graph, from which we construct a quad-dominant mesh whose edges follow the optimized direction fields. For de-homogenization, the widths of the edges are varied per element and along different directions according to the average direction and volume fraction of the optimized cells covered by an element. 

We draw inspiration from prior work on structural design using principal stress lines (PSLs)~\cite{kratz2014tensor,kwok2016structural,WWW2022stress}. The structures following principal stress directions are continuous, and this regularity is often appreciated in industrial design and architecture~\cite{Loos2022SMO}. These prior explorations, however, make use of the principal stress directions in the stress field of a solid object with isotropic material. It deviates from the stress tensor field of the final optimized structure which is composed of orthotropic cells. Furthermore, the uniform sampling of the stress lines has been a challenge, and the beam width was typically assigned based on heuristics. 
For example, Kwok et al.~\cite{kwok2016structural} propose an iterative optimization process in which lattice structures along PSLs appear incrementally. This method works for concentrated loads but it is challenging to cope with distributed loads on the design domain. Wang et al.~\cite{WWW2022stress} use the space-filling and evenly-spaced PSLs for structural design, where the beam width is adjusted using a strain energy-based importance metric. These approaches are attractive for their computational efficiency, yet the stiffness of the obtained structures are sub-optimal. In contrast to these works, we use the result of homogenization-based topology optimization for streamline tracing and to de-homogenize the single elements in the resulting mesh structure. We show that this creates structures with significantly improved stiffness.

The remainder of this paper is organized as follows. We first give an overview of the proposed method in Section~\ref{sec:overview}. In Section~\ref{sec:HomoOpti}, we review the problem formulation of the homogenization-based topology optimization. In Section~\ref{sec:param} we describe the construction process of a space-filling mesh from the direction fields that are optimized via homogenization. Mesh-based  de-homogenization is presented in Section~\ref{sec:deHomo}, and we demonstrate the effectiveness of our approach in a variety of examples in Section~\ref{sec:results}. Section~\ref{sec:conclusion} concludes the paper with a discussion of the proposed approach as well as future research directions. 

\begin{figure*}
     \centering
     \includegraphics[width=0.95\linewidth, trim=0.0cm 0.0cm 0.0cm 0.0cm, clip=true]{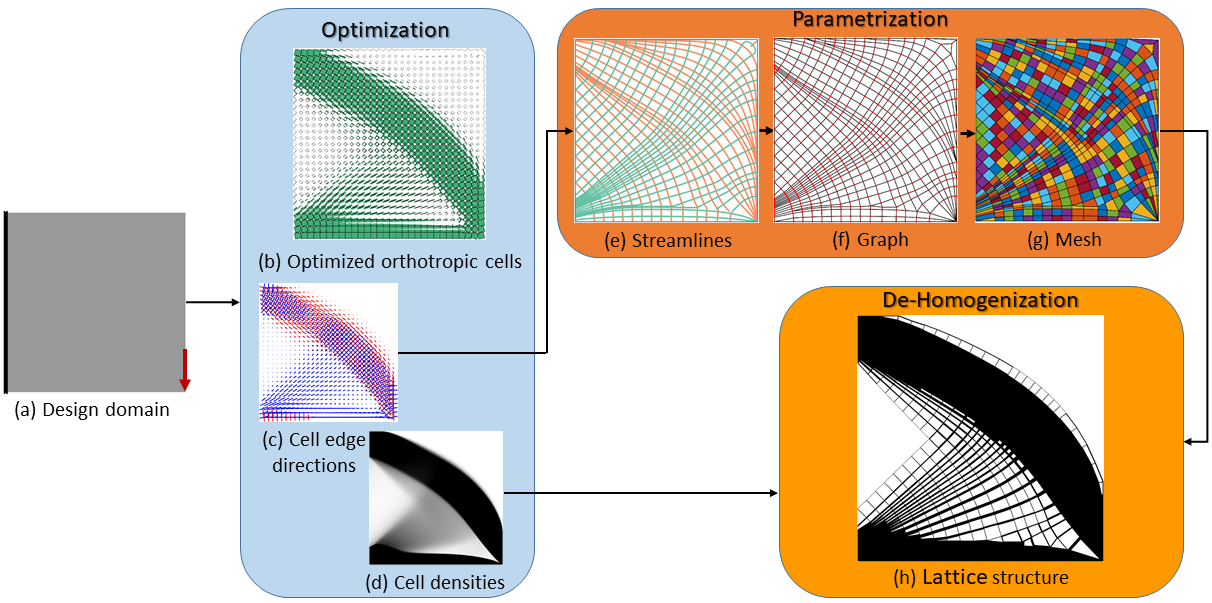}
     \caption{Method overview. (a) The design domain and boundary conditions. (b) The optimized distribution of orthotropic cells from homogenization-based topology optimization. (c) The mutually orthogonal direction fields defined by the axes of the orthotropic cells. (d) The equivalent density distribution of the orthotropic cells. (e) Streamlines traced along the two orthogonal direction fields. (f) The graph structure extracted from the streamlines. (g) The quad-dominant mesh obtained from the graph structure. (h) The final structural design.
     }
     \label{fig:overview} 
\end{figure*}

\vspace{-4mm}
\section{Method Overview} \label{sec:overview}

Our approach comprises three major stages, which are illustrated in Fig.~\ref{fig:overview}. The input is a design domain including boundary conditions, i.e., the fixations of the domain and the external forces (Fig.~\ref{fig:overview}a). Furthermore, the material properties and the volume fraction that can be consumed by the optimized layout are set by the user. 

\vspace{-2mm}
\paragraph{\textbf{Optimization.}} In the first stage, homogenization-based topology optimization is used to optimize the orthotropic cell distribution (Fig.~\ref{fig:overview}b). From this distribution, the direction fields (Fig.~\ref{fig:overview}c) and the density distribution (Fig.~\ref{fig:overview}d) are extracted. The direction fields locally coincide with the edges of orthotropic square cells whose deposition ratio and orientation are optimized. The size of the rectangular hole within each cell determines the local material consumption, and the ratio between the widths of the cell's edges determine the local material anisotropy.

\vspace{-2mm}
\paragraph{\textbf{Parametrization.}} In the second stage, first a domain-filling and evenly-spaced set of streamlines is computed in the direction fields (Fig.~\ref{fig:overview}e). Then, a graph structure is constructed, in which adjacent streamline intersection points and intersection points with the initial domain boundaries are connecting via edges (Fig.~\ref{fig:overview}f). The graph is finally converted into a mesh that is composed of mostly quadrilateral and few triangular cells bounded by the edges of the graph (Fig.~\ref{fig:overview}g).

\vspace{-2mm}
\paragraph{\textbf{De-homogenization.}} In this last stage, the final structural design (Fig.~\ref{fig:overview}h) is computed by jointly using the quad-dominant mesh, the optimized density distribution, and the anisotropy of optimized square cells. The mesh structure divides the design domain into a space-filling set of elements whose interior is filled with material according to the optimized density distribution and the anisotropy of each element.

\vspace{-4mm}
\section{Homogenization-based Topology Optimization} \label{sec:HomoOpti}

\begin{figure}
     \centering
     \includegraphics[width=0.98\linewidth, trim=0.0cm 0.0cm 0.0cm 0.0cm, clip=true]{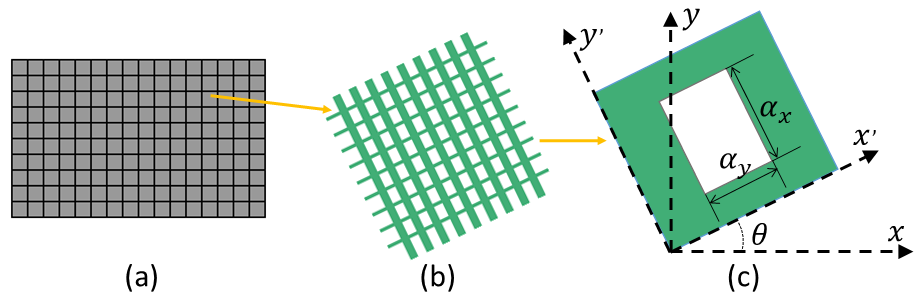}
     \caption{(a) The design domain is discretized into bilinear square grids. (b) Each square element is assumed to be filled by the orthotropic material. (c) The size and orientation of the approximately equivalent orthotropic cell, i.e., the square element with rectangular hole, are taken as design variables in homogenization-based topology optimization.}
     \label{fig:unitCellSchematic} 
\end{figure}

For structures under a single load, the theoretically optimal structural layout can be approximated by optimizing the distribution of square cells with a rectangular hole~\cite{Bendsoe1988CMAME}. As illustrated in Fig.~\ref{fig:unitCellSchematic}a, the design domain is discretized into finite elements. Each element represents a repetition of an adapted configuration of the unit cell. The square cell has a unit side length. Within it, there is a rectangular hole (Fig.~\ref{fig:unitCellSchematic}c). The configuration of the unit cell is thus described by the hole sizes $\alpha_x$ and $\alpha_y$ and rotation angle $\theta$. The mechanical properties of the unit cell is orthotropic. In this paper we refer to these adapted cells as \textit{orthotropic cells}. The density or deposition ratio ($\rho_e$) of each cell is measured by $1-\alpha_x \alpha_y$. The elasticity tensor of the orthotropic cell is computed by
\begin{equation} \label{eqn:elasticityTensor}
    C(\alpha_x, \alpha_y, \theta) = R^T(\theta) C^H(\alpha_x, \alpha_y) R(\theta),
\end{equation}
where $R(\theta)$ is the rotation matrix, and $C^H(\alpha_x, \alpha_y)$ represents the effective elasticity tensor for an axis-aligned unit cell with $\alpha_x, \alpha_y$, evaluated by numerical homogenization.

The structural design is formulated as compliance minimization,
\begin{align}
    \displaystyle \min \limits_{\bm{\alpha}_x,\:\bm{\alpha}_y,\:\bm{\theta}} \quad & \frac{1}{2} \bm{F}^{T} \bm{U}, \label{eqn:obj}\\
    \displaystyle \mathrm{subject\,to}  \quad & \bm{K}(\bm{\alpha}_x,\:\bm{\alpha}_y,\:\bm{\theta}) \bm{U} = \bm{F}, \label{eqn:FEA}\\
    \displaystyle & \frac{1}{n} \sum\limits_e \rho_e - \alpha_{global} \leq 0, \label{eqn:cons}\\
    \displaystyle & 0 \leqslant \alpha_x,\:\alpha_y \leqslant 1. \label{eqn:DD}
\end{align}
Here the objective is to minimize the elastic energy. $\bm{F}$ is the loading vector. $\bm{U}$ is the displacement vector, obtained by solving the static equilibrium equation (Eq.~\ref{eqn:FEA}). $\bm{K}$ is the stiffness matrix in finite element analysis. $n$ is the number of finite elements. $\alpha_{global}$ is the volume fraction prescribed by the user.

We use the procedure reported by Groen and Sigmund~\cite{Groen2018IJNME} for solving the optimization problem. The educational code for this was provided in the review article~\cite{Wu2021SMO}. In this procedure, $\alpha_x$ and $\alpha_y$ are optimized by gradient-based numerical optimization, while the rotation angle ($\theta$) in each iteration is determined by the corresponding principal stress direction. 

\begin{figure*}
     \centering
     \includegraphics[width=0.98\linewidth, trim=0.0cm 0.0cm 0.0cm 0.0cm, clip=true]{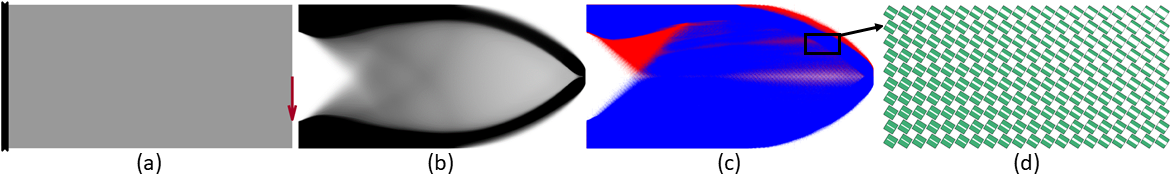}
     \caption{(a) The design domain. (b), (c) The optimal density layout and the corresponding direction field produced by the homogenization-based optimization. (d) Closeup of the optimized orthotropic cells sampled from the highlighted region in (c).
     }
     \label{fig:HomoMethod} 
\end{figure*}

Figure~\ref{fig:HomoMethod} demonstrates the results of the homogenization-based optimization for the ``Cantilever'' model, showing the initial domain and external forces (a), the extracted density layout (b) and the direction fields of the optimized orthotropic cell distribution (c). Fig.~\ref{fig:HomoMethod}d provides a closeup view of the layout of the orthotropic cells. This layout is not directly manufacturable, and needs to be transformed into a consistent geometry.

\vspace{-4mm}
\section{Parametrization} \label{sec:param}

The final goal of our approach is to convert the locally spatial-varying orthotropic cells into a globally consistent geometry. While thick sub-structures or one single solid block should be placed in dense regions, in less dense regions only few thin sub-structures are required. These sub-structures follow the optimized direction fields. In contrast to previous approaches that find a fine-grid scalar field with constraints on its gradient, we trace streamlines along the optimized direction fields. This ensures a global consistency of the sub-structures, and their alignment with the optimized direction fields.

\begin{figure*}
     \centering
     \includegraphics[width=0.98\linewidth, trim=0.0cm 0.0cm 0.0cm 0.0cm, clip=true]{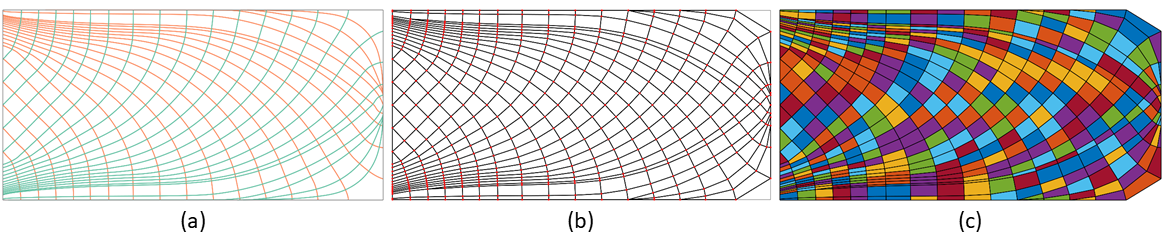}
     \caption{Parametrization by streamlines. (a) Space-filling and evenly-spaced streamlines. (b) Graph structure. (c) Quad-dominant mesh.}
     \label{fig:Parametrization} 
\end{figure*}

We use the direction fields that are optimized via the homogenization-based approach and generate a trajectory-based parametrization of the design domain.
Therefore, a uniformly distributed set of streamlines in the direction fields is computed, based on streamline computation techniques that were initially developed for visualizing 3D stress tensor fields~\cite{wang2022TSV3D}. The tool enables to compute and visualize a space-filling and evenly-spaced set of streamlines in three mutually orthogonal direction fields. It can be used to work with 2D fields (i.e., the $u$- and $v$-field in our current application), by setting the third vector field to zero. We will subsequently call the corresponding streamlines in the $u$- and $v$-field the $u$-streamlines and $v$-streamlines, respectively. The streamline construction process ensures that around each streamline an empty band is generated from which no streamline is seeded, and new streamlines are always seeded from points on existing streamlines. In this way, a fairly uniform and space-filling set of streamlines is computed (see Fig.~\ref{fig:Parametrization}a). 

Each streamline can be converted into a polyline consisting of a set of intersection points and linear connections between them ~\cite{wang2022stress}. From this representation, a graph structure with the nodes and edges, respectively, being the intersection points and piecewise linear connections between them can be easily constructed. By connecting adjacent integration points on the domain boundaries, the final graph ---due to the mutual orthogonality of the $u$- and $v$- streamlines---comprises mostly regions that are bounded by exactly four edges. Only at degenerate points and at points lying on a boundary, regions that are bounded by three edges can occur. The result of this process is shown in Fig.~\ref{fig:Parametrization}b.

Finally, the graph structure is used to discretize the design domain into a set of independent elements, i.e., the interior regions of the graph structure, so that each element can be de-homogenized independently. The orientation of the elements is given by the streamline skeleton, and the de-homogenization process proceeds by filling the elements with material according to the optimized density field. This is performed by extruding material from the edges of each element inward, according to the volume fraction of the continuous density field in each element. To do so, the graph structure first needs to be converted into an explicit cell-based mesh structure.

Since the 2D graph structure represents the connectivity (i.e. the edges) between the coordinates of the streamline intersection points, a quad-dominant mesh can be constructed in a straight forward way from this structure. By iteratively processing the local vertex and edges topology along streamlines, a mesh comprising quadrilateral and triangular elements can be computed, along with the cell topology that represents the cell adjacency information. Note that the local ordering of nodes of each quadrilateral and triangular cell needs to be consistent, i.e., either clockwise or counter-clockwise. Figure~\ref{fig:Parametrization}c shows the constructed mesh from the graph structure in Fig.~\ref{fig:Parametrization}b. 

\paragraph{Singularities.}
To obtain a consistent mesh structure from the streamline skeleton, singularities in the direction fields need to be determined and treated in a special way. In our case, where the direction fields coincide with principal stress directions, singularities occur at so-called degenerate points of the corresponding stress field, i.e., points where the two eigenvalues of the stress tensor in the underlying stress tensor field become indistinguishable. In the seminal work by Delmarcelle and Hesselink ~\cite{delmarcelle1994topology} both the classification of degenerate points and their numerical computation is discussed. In the vicinity of degenerate points a set of hyperbolic and parabolic sectors exist, in which similar patterns of neighboring trajectories are observed. The topological skeleton consists of the boundaries between adjacent sectors---so-called separatrices---and indicate pathways along which the forces are steered towards the degenerate points. By first extracting the degenerate points and computing the topological skeleton, the separatrices can then considered as seed streamlines as described before, so that an evenly-spaced set of streamlines is computed in each sector. Let us refer to the work by Wang et al.~\cite{wang2022stress} for a more detailed description of the implemented procedure. Figure~\ref{fig:singularity}b and~\ref{fig:LshapeMBB}f show the embedding of singularities into the computed streamlines.

\vspace{-4mm}
\section{De-Homogenization} \label{sec:deHomo}

\begin{figure}
     \centering
     \includegraphics[width=0.98\linewidth, trim=0.0cm 0.0cm 0.0cm 0.0cm, clip=true]{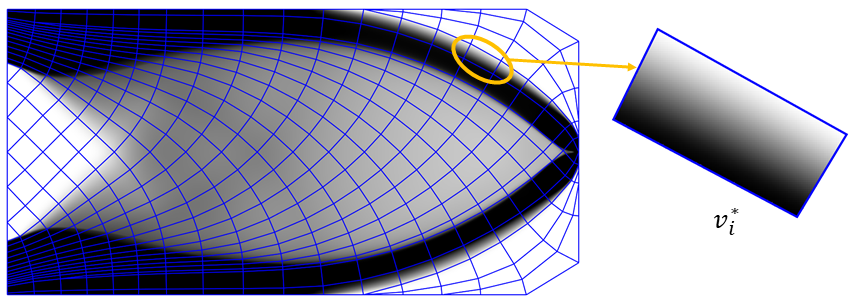}
     \caption{Correspondence between continuous material field and mesh elements, facilitating the assignment of a material budget $v_i^*$ to each mesh element and de-homogenization of each element separately.
     }
     \label{fig:deHomo} 
\end{figure}

In order to de-homogenize the optimal density layout, i.e., to convert the continuous density layout into a binary one, we utilize the constructed quad-dominant mesh and de-homogenize the region covered by each mesh element separately. As shown in Fig.~\ref{fig:deHomo}, each element covers a certain region in the domain. The material in each region, i.e., the deposition ratio, should be re-distributed so that a) a binary material layout is generated, b) a continuous transition at the element boundaries is obtained, and c) the orthotropic cells' orientations, which have been optimized with respect to the object's compliance, is reflected in the binary material layout. The target deposition ratio $v_i^*$ of a mesh element (see Fig.~\ref{fig:deHomo}) is measured by $\frac{D}{M}$, where $M$ is the number of orthotropic cells located in the region covered by the element, and $D$ is sum of the density values over all these cells. The de-homogenized mesh element should keep this deposition ratio after de-homogenization.

Our approach allows to easily achieve different granularity levels of the streamlines by setting the seeding rate. Since the resolution of the corresponding quad-dominant mesh varies spatially and is not necessarily at a resolution comparable to the finite element discretization used in topology optimization, the density and directions need to be resampled from the finite element grid to compute the target deposition ratio of a mesh element. This is performed via bi-linear interpolation at a set of sampling points in each mesh element.

\vspace{-4mm}
\subsection{Anisotropic mesh element} \label{subsec:aniEle}

To distribute the material in a mesh element according to the mentioned requirements, we propose to extrude the available material from the edges of each element inward. By starting with a minimal edge thickness, the edges are iteratively thickened until all available material is used. In this way, we enforce a layout that aligns with the element orientation, seamlessly connects adjacent elements, and can furthermore account for an anisotropic stress distribution by adapting the edge thickness according to the mechanical properties of each element. 

The process starts by removing mesh elements which have a very low target deposition ratio (e.g., $v^{*}<0.05$), in order to avoid the generation of very thin mesh edges that can cause difficulties in the manufacturing process. Similarly, mesh elements with a large deposition ratio (e.g., $\frac{D}{M}>0.95$) are made fully solid. The edges of all remaining mesh elements are set to a minimum thickness $t_0$. 
\begin{figure}
     \centering
     \includegraphics[width=0.9\linewidth, trim=0.0cm 0.0cm 0.0cm 0.0cm, clip=true]{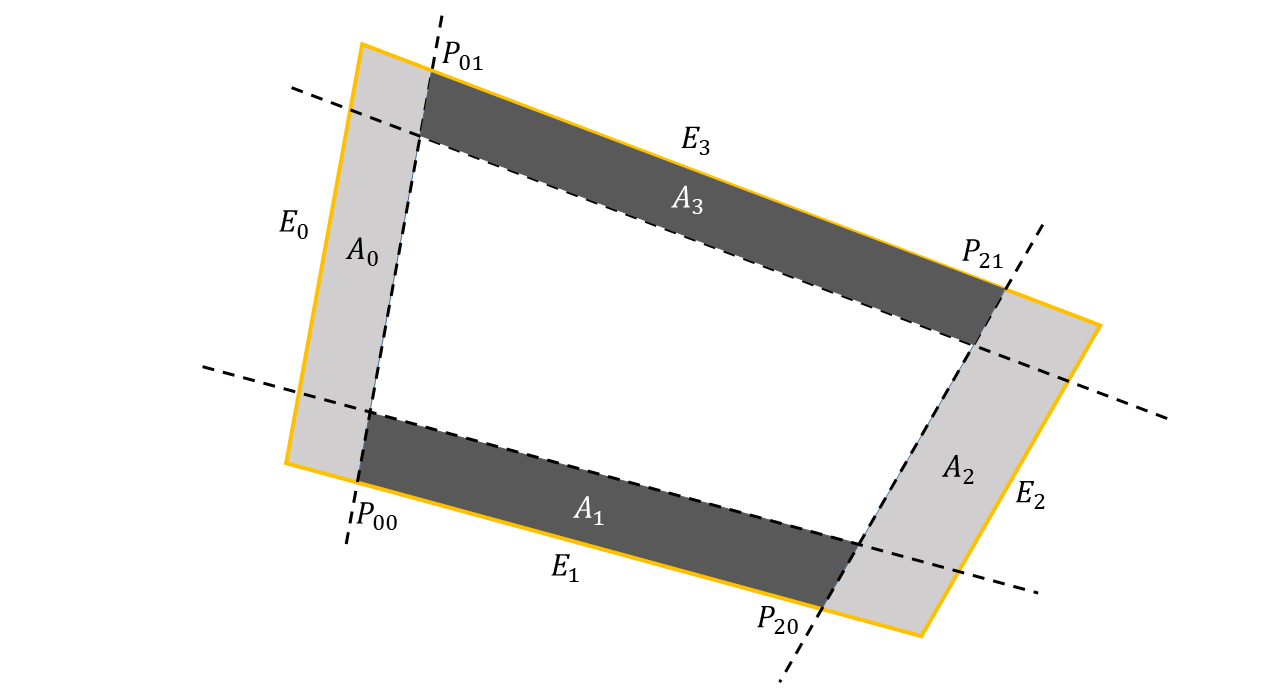}
     \caption{Schematics of computing the area subtended by edges of a certain thickness. From intersection points between dashed lines (extruded edges) and edges of the mesh element the areas can be computed.}
     \label{fig:edgearea} 
\end{figure}

Taking a single mesh element as shown in Fig.~\ref{fig:edgearea}, its deposition ratio $v$ is computed as the sum of the areas $A_i$ covered by each extruded edge, subtracting the sub-areas that are counted twice, and then dividing by the total area of the element. We start with two opposite edges and compute for each edge the intersection points between the extruded edge and the respective other mesh edges ($P_{00},P_{01},P_{20},P_{21}$ in Fig.~\ref{fig:edgearea}). Including the endpoints of the mesh edges, this gives two quadrilaterals $A_0, \:A_2$, whose areas can be computed via triangulation. For the other two mesh edges, we use the newly computed edge intersection points and the intersection points between the four extruded edges, and compute two quadrilaterals representing the missing areas $A_1, A_3$. Now, the thickness $t$ of each edge can be increased iteratively from the initial value $t_0$, until the actual deposition ratio $v$ approaches the target $v^{*}$, i.e., the available material budget is used. 

\begin{figure}
     \centering
     \includegraphics[width=0.98\linewidth, trim=0.0cm 0.0cm 0.0cm 0.0cm, clip=true]{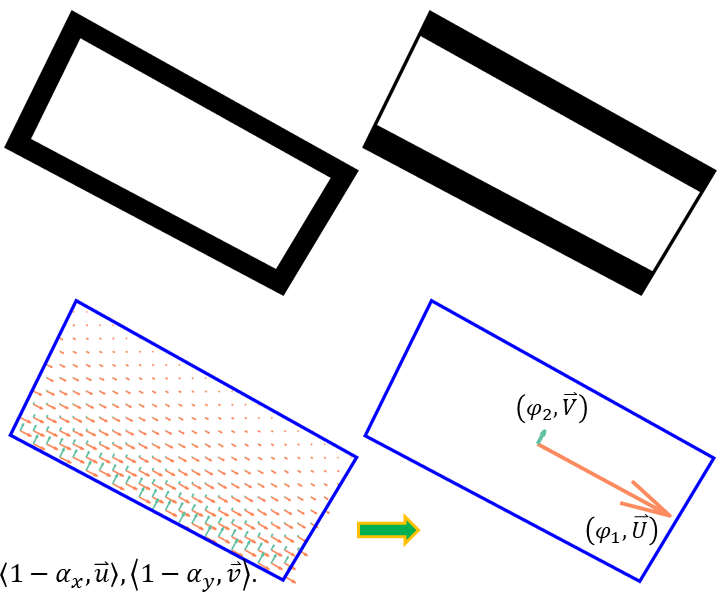}
     \caption{Top: the edges are equally thickened (left) or thickened (using the same amount of material) according to the edge thicknesses of the orthotropic cells they cover (bottom left). Bottom: For the set of orthotropic cells covered by a mesh element, representative edge thicknesses and orientation are computed via averaging. The arrow length indicates the thickness of edges along the pointing direction. 
     }
     \label{fig:latEle} 
\end{figure}

In order to match the mechanical properties of the set of orthotropic cells covered by a single mesh element, we start with a minimal edge thickness, and then thicken the edges according to the ratio of the edge thicknesses of the orthotropic cells (Fig.~\ref{fig:latEle}). The edge thicknesses $1-\alpha_x$ and $1-\alpha_y$ as well as the orientation of an orthotropic cell have been optimized to maximize stiffness of the resulting layout. As such, if all cells covered by a mesh element have the same thickness ratio, and are consistently orientated with the mesh element, the material should be deposited along the element edges so that the thickness ratio of the cells is maintained. However, since the ratio and orientation vary across the cells, in general, we first need to compute representative values for both (see Figure~\ref{fig:latEle} bottom for an illustration). 

\begin{figure*}[h]
     \centering
     \includegraphics[width=0.98\linewidth, trim=0.0cm 0.0cm 0.0cm 0.0cm, clip=true]{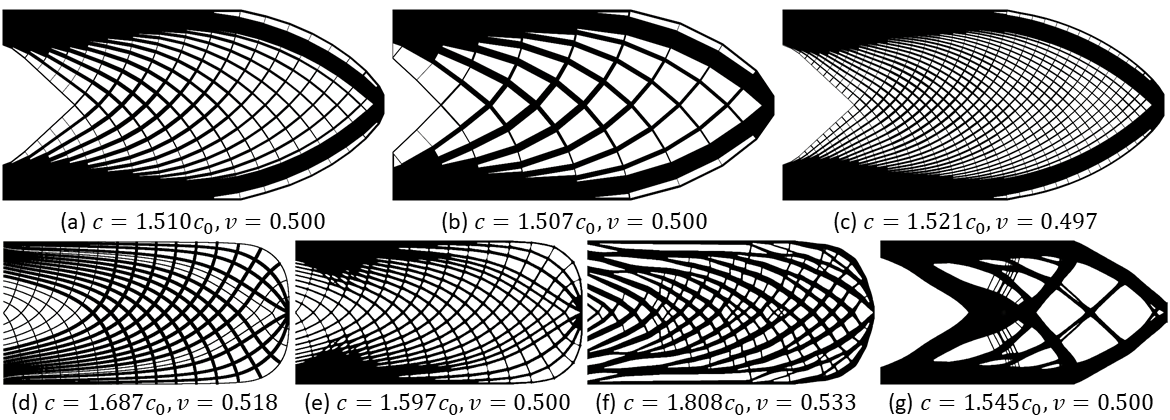}
     \caption{Top: Results with the proposed de-homogenization strategy. (a) to (c) demonstrate the changes due to different amounts of streamlines used in the parametrization stage. Bottom: Comparison to alternative approaches for generating a binary design via topology optimization. (d) PSL-guided material layout~\cite{WWW2022stress}. (e) PSLs-initialized density-based topology optimization using a global volume constraint~\cite{WWW2022stress}. (f) Porous-infill optimization with local volume constraints~\cite{Wu2018TVCG}. (g) Density-based topology optimization with a global volume constraint. $c_0$ and $c$, respectively, are the compliances of the fully solid domain and the shown results, $v$ is the volume fraction.
     }
     \label{fig:benchmarkExp} 
\end{figure*}

We do so by first computing the normalized thickness of the cell edges $\varphi_1$ and $\varphi_2$ following the $u$- and $v$-field, respectively, by adding up the values of the per-cell thicknesses and dividing through the maximum of the resulting values, i.e., 
\begin{equation} \label{eqn:dirFieldConvertion}
    \begin{array}{l} 
        \varphi_1^* = \sum{(1-\alpha^i_x)}, \: \varphi_2^* = \sum{(1-\alpha^i_y)}, \\ \\
        \varphi_1 = \varphi_1^* / \max(\varphi_1^*, \: \varphi_2^*), \: \varphi_2 = \varphi_2^* / \max(\varphi_1^*, \: \varphi_2^*). 
    \end{array}  
\end{equation}

Then, to determine which edge of the mesh element corresponds to $\varphi_1$ and which to $\varphi_2$, we compute the 
average direction vectors $\overrightarrow{U}$ and $\overrightarrow{V}$ of all the per-cell direction vectors $\overrightarrow{u_i}$ and $\overrightarrow{v_i}$, i.e., $\overrightarrow{U} = \mathrm{norm}(\sum \overrightarrow{u_i})$ and  $\overrightarrow{V} = \mathrm{norm}(\sum \overrightarrow{v_i})$. We let the mesh edges correspond to $\overrightarrow{U}$ or $\overrightarrow{V}$ to which they have the least directional deviation.

Now, we can introduce for each mesh edge $e_j$ a scaling factor $w_j$, which is calculated by $e_j=t_0 + w_{j}\delta$. Here $\delta$ is an increment used for adjusting the edge thickness iteratively. 
With the thicknesses and directions $(\varphi_1, \: \overrightarrow{U})$ and $(\varphi_2, \: \overrightarrow{V})$, the corresponding weighting factor $w_{j}$ of the j-$th$ element edge  is given by
\begin{equation} \label{eqn:weightingFact}
    w_{j} = \theta_{j1} < \theta_{j2} \: ? \: \varphi_1 : \varphi_2.
\end{equation}

Here, $\theta_{j1}$ and $\theta_{j2}$ are the included angles between the j-$th$ element edge and the directions $\overrightarrow{U}$ and $\overrightarrow{V}$, respectively, and we consistently use the same thickness for opposite edges in each quadrilateral element, and the triangular elements are treated as degenerate quadrilaterals where one of the edges is collapsed, specifically, for each edge, the weighting factor $w_{j}$ is determined by Eqn.~\ref{eqn:weightingFact}.

It is worth mentioning that in rare cases the per-cell directions may change considerably in a single mesh element, and thus the per-element direction becomes less representative. To cope with this, the mesh element can be subdivided into a set of smaller elements, for each of which the de-homogenization is performed as described. 

\vspace{-4mm}
\section{Results} \label{sec:results}

We demonstrate our de-homogenization approach with several examples, and compare the results to those of density-based and de-homogenization approaches. In all cases, the design domains are discretized by Cartesian grids with unit-size. The Young's modulus and Poisson's ratio are set to 1.0 and 0.3, respectively. Homogenization-based topology optimization is performed with the Matlab code provided in~\cite{Wu2021SMO}. We terminate the optimization process after 200 iterations. We have implemented the proposed parametrization and de-homogenization operations in Matlab as well. All experiments have been carried out on a desktop PC with an Intel Xeon CPU at 3.60GHz. In all of our experiments, the time for parametrization and de-homogenization is less than a minute.

\paragraph{\textbf{Comparison to density-based approaches.}}
In our first experiment, we use the cantilever model described in Fig.~\ref{fig:HomoMethod}a to demonstrate the properties of the proposed de-homogenization approach and compare the results to those of density-based topology optimization.

Figure~\ref{fig:benchmarkExp} (top) shows the de-homogenization results for different streamline densities, resulting in an increased or decreased number of ever smaller or larger mesh elements, respectively. The compliances of the designs with different granularity vary only slightly. With $c_0$ being the compliance of the fully solid domain, one can see that with the same amount of material all resulting designs achieve almost the same compliance of roughly $1.5c_0$. The compliance of the de-homogenized binary layouts is slightly higher than that from homogenization-based optimization ($1.447c_0$). 

In Fig.~\ref{fig:benchmarkExp} (bottom), we compare our results to those generated by the stress trajectory-guided structural design by Wang et al.~\cite{WWW2022stress}, the porous infill approach using local volume constraints by Wu et al.~\cite{Wu2018TVCG}, and density-based topology optimization with a global volume constraint. In all examples, the same number of simulation element as for de-homogenization in our approach is used. 

Fig.~\ref{fig:benchmarkExp}d,e show the results of stress trajectory-guided structural design. In (d), the material is distributed along principal stress trajectories of the solid object under load, and the thickness of the material is adapted according to the accumulated strain energy along each trajectory. Fig.~\ref{fig:benchmarkExp}e shows the optimized material layout when the material field in (d) is used as initialization for topology optimization with a global volume constraint. The generated layouts also show a very regular structural design, but a considerably higher compliance than the de-homogenization approach.
The latter can also be observed when comparing to porous infill optimization (Fig.~\ref{fig:benchmarkExp}f) which applies a strict constraint on local volume. Here, besides having a smaller compliance, the de-homogenized layout (Fig.~\ref{fig:benchmarkExp}a-c) also shows a more regular structural layout. Notably, while porous infill optimization generates many bifurcations, i.e., solid strands that merge and split, the de-homogenization approach, per construction, results in a grid-like structure mostly comprising quadrilateral elements. Finally, compared to density-based topology optimization with a global volume constraint (Fig.~\ref{fig:benchmarkExp}g), our result still shows a slightly smaller compliance, yet the results are far more regular and, are expected to exhibit higher stability when the load conditions are changed or certain parts undergo damage, as demonstrated for evenly-spaced, space-filling structures in~\cite{Wu2018TVCG,wang2022stress}.

\begin{figure*}
     \centering
     \includegraphics[width=0.98\linewidth, trim=0.0cm 0.0cm 0.0cm 0.0cm, clip=true]{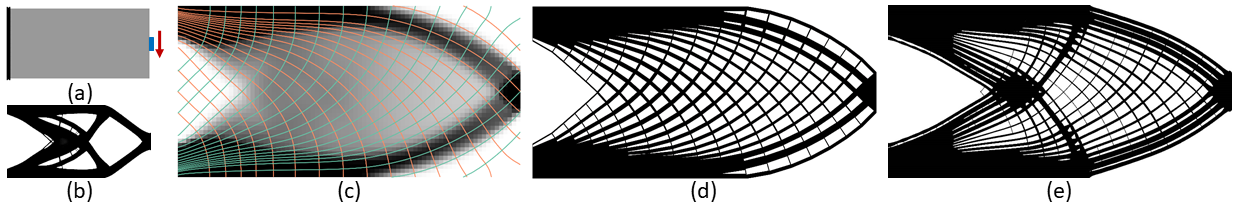}
     \caption{Comparison of de-homogenization approaches on a cantilever beam. (a) The design domain and boundary conditions. (b) The result of density-based topology optimization with a global volume constraint simulated ($c=62.56$, $v=0.500$). (c) Optimal density layout, superimposed with streamlines. (d) The de-homogenized structural design by our method ($c=60.04$, $v=0.500$). (e) The result of the projection-based de-homogenization ($c=58.57$, $v=0.510$). The image is reproduced from~\cite{Groen2018IJNME}.}     
     \label{fig:comparison2IJNME2018} 
\end{figure*}

\begin{figure*}
     \centering
     \includegraphics[width=0.98\linewidth, trim=0.0cm 0.0cm 0.0cm 0.0cm, clip=true]{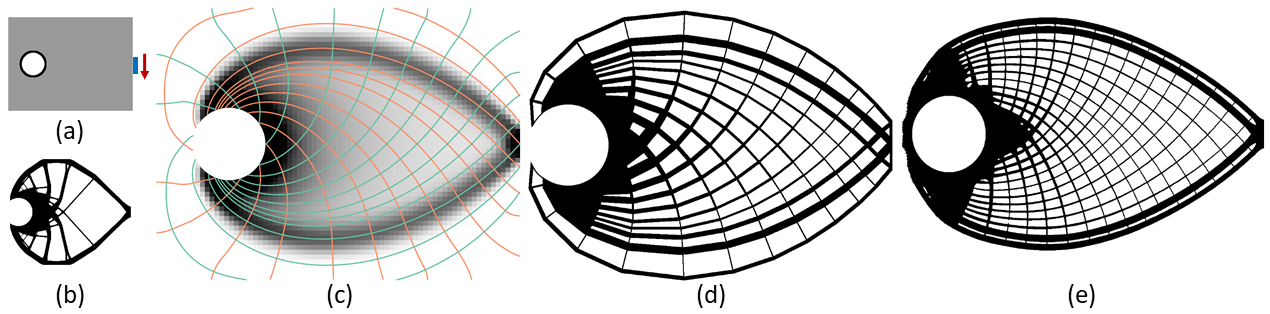}
     \caption{Comparison of de-homogenization approaches on a Michell's structure. (a) The design domain and boundary conditions. (b) The result of density-based topology optimization with a global volume constraint ($c=64.452$, $v=0.250$). (c) Optimal density layout, superimposed with streamlines. (d) The de-homogenization using our approach ($c=62.301$, $v=0.250$). (e) The result of the projection-based de-homogenization ($c=67.830$, $v=0.252$). The image is reproduced from~\cite{Groen2018IJNME}.}
     \label{fig:MichellProblem} 
\end{figure*}

\begin{figure*}
     \centering
     \includegraphics[width=0.98\linewidth, trim=0.0cm 0.0cm 0.0cm 0.0cm, clip=true]{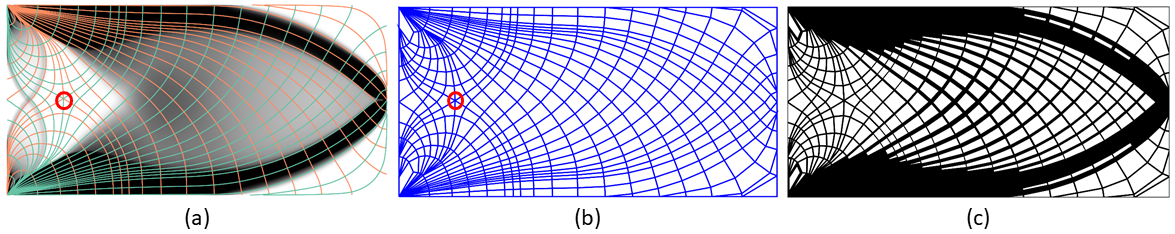}
     \caption{Tests on a cantilever beam fixed by the endpoints of its left boundary. (a) Optimal density layout ($c^*=1.138c_0$, $v=0.500$) and streamlines. (b) The generated streamline graph. The degenerate point (singularity) is marked by the red circle. (c) The de-homogenized structural design ($c=1.170c_0$, $v=0.500$).}
     \label{fig:singularity} 
\end{figure*}

\begin{figure*}
     \centering
     \includegraphics[width=0.98\linewidth, trim=0.0cm 0.0cm 0.0cm 0.0cm, clip=true]{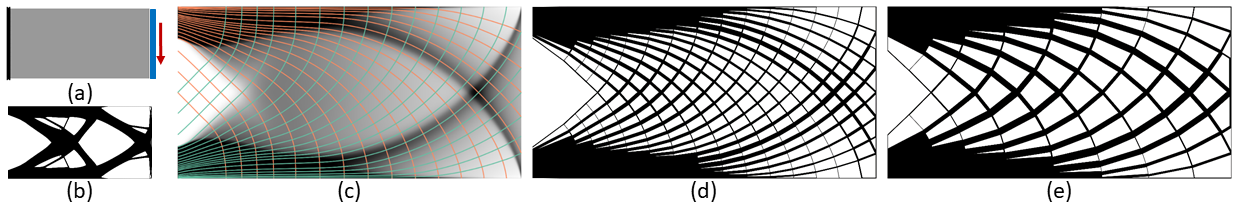}
     \caption{(a) A cantilever under a distributed load along the right edge. (b) Result of density-based topology optimization with a global volume constraint ($c=1.675c_0$, $v=0.500$). (c) Optimal density layout ($c^*=1.533c_0$, $v=0.500$) and streamlines. (d) De-homogenized structural design ($c=1.661c_0$, $v=0.500$). (e) De-homogenized structural design with less streamlines ($c=1.680c_0$, $v=0.500$).}
     \label{fig:distributedForce} 
\end{figure*}

\paragraph{\textbf{Comparison to de-homogenization approaches.}} In Fig.~\ref{fig:comparison2IJNME2018}, we compare the result that is obtained with our approach to the result of projection-based de-homogenization by Groen and Sigmund~\cite{Groen2018IJNME}. To  match their model configuration, the force applied to the cantilever model has been changed accordingly, and homogenization-based topology optimization is performed at a coarse grid resolution of $100\times50$. Density based topology optimization using a global volume constraint is also included here as a reference. It is computed at a grid resolution of $1600\times800$. It can be seen that the compliance of the layout generated by our approach (d) is only slightly higher than that of the layout produced by the method of Groen and Sigmund in (e), yet using a little less material. The layout in (e) has some concentrated clusters in the middle of the domain, while our approach generates a more uniform grid-like material layout. This difference is likely due to the parameter setting (e.g., the filter size) in the homogenization-based topology optimization.
While Groen et al. perform an optimization to generate a consistent binary pattern at finer resolution ($1600\times800$ in the example) from the coarse grid results, in our method the resolution of the final layout is controlled by the density of seeded streamlines. Since streamlines are always traced in the initial domain, they always stay entirely within the domain. The optimized quantities required to trace  streamlines and de-homogenize the final mesh elements are reconstructed via bilinear interpolation from the coarse grid. 

Figure~\ref{fig:MichellProblem} shows the structural design that is generated by our method when applied to the Michell's structure according to the specification in~\cite{Groen2018IJNME}. The coarse and fine grid resolutions used for optimization and de-homogenization are $80\times60$ and $1280\times960$, respectively. In this case, the compliance of our design (Fig.~\ref{fig:MichellProblem}d) is lower than that of the projection-based de-homogenization (Fig.~\ref{fig:MichellProblem}e). As a side note, perfect symmetry is not achieved because the streamline seeding process is not designed to consider symmetry in the design domain or the underlying direction field. If symmetry is known beforehand, however, the seeding process can be easily adapted to consider it.

From these two examples, it can be found that the compliance from the proposed streamline based approach is comparable to that from the projection-based approach that requires solving for a scalar field by optimization. On this aspect, the computation involved in our approach is more efficient. For example, the parameterization and de-homogenization for Fig.~\ref{fig:comparison2IJNME2018}d took about 40 seconds, while from ~\cite{Groen2018IJNME}, the de-homogenization for Fig.~\ref{fig:comparison2IJNME2018}e took more than 2 minutes.

\paragraph{\textbf{Singularity treatment.}}
As we described in Section~\ref{sec:param}, our proposed approach can handle situations where singularities exist in the direction fields that are obtained via homogenization-based optimization. Such singularities usually incur discontinuities during streamline tracing, and they furthermore result in low convergence for density-based topology optimization under local volume constraints~\cite{wang2022stress}. The singularity can be detected by topology analysis of the orthogonal direction fields. As an example, we again use the cantilever model (Fig.~\ref{fig:HomoMethod}a), but now replace the distributed fixation condition with point fixations applied on the endpoints of the left boundary. Fig.~\ref{fig:singularity}a highlights a singularity in the left part of the domain, where 3 $u$-streamlines and 3 $v$-streamlines converge to a single point. This type of singularity is termed a trisector degenerate point in stress topology analysis, and the 6 streamlines are the corresponding topological skeleton. Figure~\ref{fig:singularity}b shows the generated mesh, which demonstrates that a consistent structure can be obtained around the singularity. The de-homogenized result is shown in Fig.~\ref{fig:singularity}c. 

\paragraph{\textbf{Distributed loads.}}
Our de-homogenization approach naturally works well also for distributed loads. Fig.~\ref{fig:distributedForce}a shows the structural design problem under distributed loads. Fig.~\ref{fig:distributedForce} compares the results of density-based topology optimization with a global volume constraint to those of our proposed de-homogenization method using different streamline densities. The compliance from density-based topology optimization is between the tight range of compliances of de-homogenized structures with two different streamline densities. The deviation of the compliances of the de-homogenized structures (d and e) from the compliance in homogenization (c) is less than $10\%$. 

\begin{figure}
     \centering
     \includegraphics[width=0.98\linewidth, trim=0.0cm 0.0cm 0.0cm 0.0cm, clip=true]{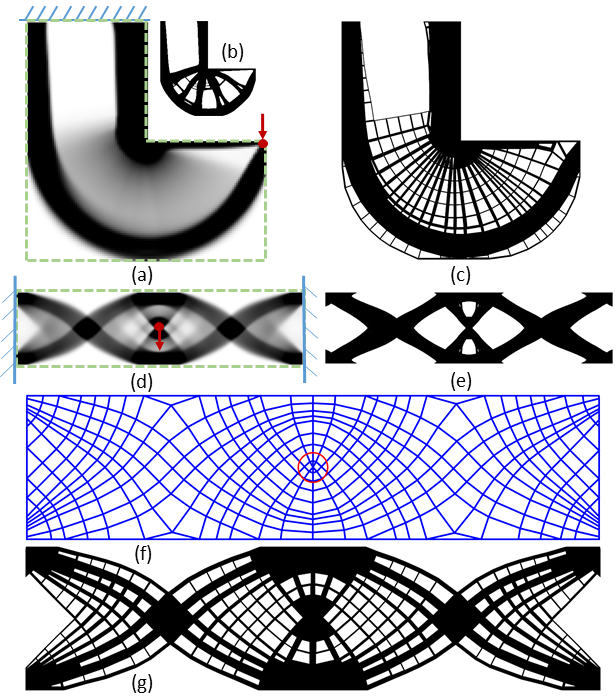}
     \caption{(a) Optimal density layout ($c^*=1.332c_0$, $v=0.500$) for an L-shaped beam under a point load. (b) Inlet shows the binary result of density-based topology optimization with a global volume constraint ($c=1.399c_0$, $v=0.500$). (c) De-homogenization result ($c=1.397c_0$, $v=0.500$). (d) Optimal density layout ($c^*=1.663c_0$, $v=0.500$) for a double-clamped beam. (e) Binary result of density-based topology optimization with a global volume constraint ($c=1.808c_0$, $v=0.500$). (f) Streamline graph used for de-homogenization. A degenerate point is marked by a red circle. (g) De-homogenization result ($c=1.763c_0$, $v=0.500$).
     }
     \label{fig:LshapeMBB} 
\end{figure}

\paragraph{\textbf{L-shape panel and MBB beam.}}
We have also tested our approach on an L-shaped beam and a double-clamped beam. Figure~\ref{fig:LshapeMBB} shows the optimized results. In both cases, the compliance of the de-homogenized layout is about $5\%$ higher than the compliance after homogenization. It is worth noting that also in the stress field of the double-clamped beam a degenerate point occurs, which, according to the topological skeleton, generates a grid composed of triangular and quadrilateral mesh elements around it. In these two examples, as in previous examples, the compliance of the de-homogenized structure is lower than that from the density-based approach with a global volume constraint.

\vspace{-4mm}
\section{Conclusion and Future Work}\label{sec:conclusion}

In this paper, we have introduced a novel streamline-based parametrization of a design domain to de-homogenize the optimal continuous density layout produced by homogenization-based topology optimization. The compliance of the de-homogenized high-resolution structures is very close to that of the optimal design from homogenization-based optimization, and it is consistently superior to the compliance achieved via density-based topology optimization. The resulting structures exhibit a globally regular appearance, uniformly covering the domain with quad-dominant mesh elements. 

In the current work we did not strive for an efficient implementation of the method. However, streamline integration and intersection computation can be effectively parallelized, for instance, on a GPU. The intersection points are already ordered along the streamlines, and graph as well as mesh construction requires only local access operations to adjacent streamlines or intersection points. Thus, we believe that the entire approach can be implemented on the GPU so that even instant de-homogenization is possible once the continuous density layout is available. We will consider such an implementation in future work, and investigate the possibility for designers to probe different streamline densities and seeding strategies. Finally, we are particularly interested in extending this approach to design 3D beam-like lattice structures. A challenge here is that the intersection of independently traced streamlines in 3D happen only coincidentally. The optimization approach for constructing stream surfaces from~\cite{stutz2022synthesis} could be an interesting direction to further explore.

\vspace{-4mm}
\begin{acknowledgment}
This work was supported by the German Research Foundation (DFG) under grant number WE 2754/10-1.
\end{acknowledgment}

%
% ...

\bibliographystyle{asmems4}

\vspace{-5mm}
\bibliography{asme2e, _TOP}



\end{document}